\documentclass[reprint,superscriptaddress,aps,prl,twocolumn]{revtex4-1}
\usepackage[T1]{fontenc}       

\usepackage{color}
\usepackage{amsmath, amsthm, amsfonts}    
\usepackage{amssymb}
\usepackage{bm}
\usepackage{comment}
\usepackage[normalem]{ulem}
\usepackage[unicode=true,pdfusetitle, bookmarks=true,bookmarksnumbered=false,
 bookmarksopen=false,breaklinks=true,pdfborder={0 0 0},backref=false,
 colorlinks=true,citecolor=magenta]{hyperref}

\usepackage{graphicx}   
\usepackage{hyperref}   

\usepackage[dvipsnames]{xcolor}
\usepackage[makeroom]{cancel}
\usepackage[normalem]{ulem}

\renewcommand{\Re}{\operatorname{Re}}

\def\msdcm{\Delta^2_{cm}}
\def\msdop{\Delta^2_{1}}

\def\ii{\mathrm{i}}
\def\bfuno{\mathbf{F}^{(1)}}
\def\bfdos{\mathbf{F}^{(2)}}

\begin{document}


\title{Strong collective currents of gold nanoparticles in an optical vortex lattice}

\author{R. Delgado-Buscalioni}\email{rafael.delgado@uam.es}
\affiliation{Department of Theoretical Condensed Matter Physics,
  Universidad Aut\'onoma de Madrid, 28049 Madrid, Spain}
\affiliation{
  Condensed Matter Physics Center (IFIMAC), 
  Universidad Aut\'onoma de Madrid, 28049 Madrid, Spain}

\author{M. Mel\'endez}
\affiliation{Department of Theoretical Condensed Matter Physics,
  Universidad Aut\'onoma de Madrid, 28049 Madrid, Spain}
 
\author{J. Luis-Hita}
\affiliation{
  Condensed Matter Physics Center (IFIMAC), 
  Universidad Aut\'onoma de Madrid, 28049 Madrid, Spain}
  \affiliation{Donostia International Physics Center (DIPC),
  Paseo Manuel Lardizabal 4, 20018 Donostia-San Sebastian, Spain}

\author{M. I.  Marqu\'es}
  \affiliation{
  Condensed Matter Physics Center (IFIMAC), 
  Universidad Aut\'onoma de Madrid, 28049 Madrid, Spain}

\affiliation{Department of Material Physics,
  Universidad Aut\'onoma de Madrid, 28049 Madrid, Spain
  }
  
 \affiliation{Instituto Nicol\'as Cabrera,
    Universidad Aut\'onoma de Madrid, 28049 Madrid, Spain}
  
\author{J. J. S\'aenz}
\affiliation{Donostia
  International Physics Center (DIPC),
  Paseo Manuel Lardizabal 4, 20018 Donostia-San Sebastian, Spain}
\affiliation{IKERBASQUE, Basque Foundation for Science,
  48013 Bilbao, Spain}

\begin{abstract}

Gold nanoparticles moving in aqueous solution under a optical vortex lattice are 
shown to present a complex collective optofluidic dynamics. Above a critical 
field intensity and concentration the system presents a spontaneous transition 
towards synchronised motion, driven by nonconservative optical forces, thermal 
fluctuations and hydrodynamic pairing. The system exhibits a rich assortment of 
collective dynamics such as strong unidirectional currents of nanoparticles at 
speeds of centimetres per second.  This relatively simple optofluidic setup 
offers an alternative way to control mass and heat transport at the nanoscale, 
which has been so far elusive.

\end{abstract}

\date{today}
\maketitle

Efficient transport of nanoscale objects is a very active area of research where 
optofluidics is heavily involved \cite{nature2016}. Beyond Optical Tweezers 
\cite{Jones}, light forces induced by stochastic optical fields 
\cite{Douglass12,volpe14,Brugger15} can be used to to ``activate'' and control 
nanoparticle motion.  Optically active micron-sized particles interacting in a 
time-varying fluctuating speckle light field can diffuse about three times 
faster than thermally, moving at microns per second \cite{Douglass12,volpe14}. 
However, below microns, active control becomes elusive due to strong thermal 
fluctuations \cite{bechinger2016active}.

A clever setup has recently been tested \cite{nature2016} using plamon-enhanced 
optical trapping to fix nanoantennas in microscopic arrangements. In conjunction 
with an oscillating electric field, photo-induced heating of these nanoantennas 
leads to microscale natural convection of the surrounding liquid at velocities 
of tens of $\mu\mathrm{m/s}$ \cite{nature2016}.  The driving mechanisms in these 
previous examples are either of stochastic origin (random speckle patterns) 
\cite{Douglass12,volpe14} or deterministic (buoyancy) \cite{nature2016}.

Instead, {\em self-organized collective motion} \cite{Vicsek2012} is inspiring 
new designs of micron-size ``artificial swarms'' 
\cite{bechinger2016active,aleksNature,bricard2013emergence} following the 
ubiquitous examples nature offers (bacteria, ants, birds and more 
\cite{Vicsek2012,ants}). Its universality is revealed by toy-models, such the as 
Vicsek model \cite{Vicsek2012}, based on extremely simple individual-activation 
and interaction rules. Hydrodynamics provides a powerful interaction kernel 
affecting, for example, the complex rotational dynamics of particles 
\cite{brzobohaty2015complex}, often leading to swarming \cite{Vicsek2012}.  The 
``colloidal rollers'' are a recent example outside optofluidics. They consist in 
micron-sized spheres in suspension which, activated by electric 
\cite{bricard2013emergence} or magnetic \cite{aleksNature} fields, rotate near a 
surface and can self-organize into flocks with cluster-speeds of centimetres per 
second \cite{aleksNature}.

In this {\em Letter} we present a new class of optically controlled active media 
on \textit{time stationary} optical curl-force fields 
\cite{hemmerich1992radiation,Albaladejo09b,Albaladejo09a,Zapata09,Berry13,Berry15,Zapata2015} 
acting on a suspension of gold {\em nanoparticles} (NP).  In accordance with 
self-organized critical phenomena, NPs spontaneously synchronize moving 
coherently in unidirectional currents at speeds of $\mathrm{cm/s}$.  We have 
numerically investigated these optofluidic dynamics by developing accurate 
Brownian hydrodynamics solvers which resolve the optical force from the incident 
light and the many-body forces induced by the light scattered by each NP. The 
proposed setup is relatively simple and we hope that the striking results 
presented here will foster experimental work on this subject.

\begin{figure}[t]
  \includegraphics[scale=0.50]{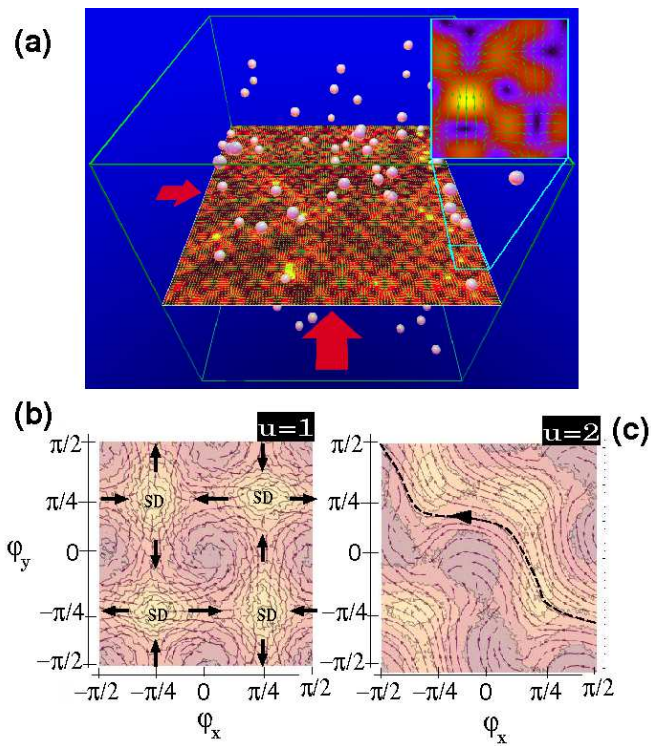}
  \caption{\label{fig1} 
    (a) Suspension of gold nanoparticles of radius $R=50\ \mathrm{nm}$ in the 
    intersection of two coherent laser beams (red arrows) with $\lambda=395\ 
    \mathrm{nm}$ polarized in the $z$ (vertical) direction and propagating in 
    the $x$ and $y$ directions (optic plane) with a $\pi/2\ \mathrm{rad}$ phase 
    lag. We show a fixed-$z$ cross section of the force field and the modulus of 
    the total electric field $|E|$ including the incident beams and the light 
    scattered by each NP (brighter regions mean higher $|E|$).  In this setup, 
    the NPs remain confined to a cubic box of side $L\simeq 7.0 \lambda$. (b) 
    Isocontours of single-particle probability density and streamlines of the NP 
    velocity field in a periodic domain, folded into the Bravais unit cell of 
    the primary optic field $x \in [-\lambda/2,\lambda/2]$ (same for $y$) 
    against phase-coordinates $\varphi_x =2\pi (x/\lambda)$. The non-dimensional 
    laser energy (see text) corresponds to $u=1$ (below) $u=2$ (above) the 
    transition to coherent NP motion (volume fraction $\phi=3.0\times 10^{-3}$).  
    Brighter regions correspond to denser domains. The unstable saddle nodes 
    (SD) of the primary force field (Eq. \ref{eq:force_field}) are indicated in 
    the $u = 1$ panel.  The dashed line for $u=2$ illustrate a zig-zag path 
    followed by NPs under coherent dynamics.}
\end{figure}

We consider a suspension of $N$ gold NPs of $R= 50\ \mathrm{nm}$ radius in water 
at dilute volume fraction ($\phi = 4\pi R^3 N/(3 V) \sim 10^{-4}$) exposed to an 
optical field formed by the intersection of two perpendicular coherent laser 
beams \cite{hemmerich1992radiation,Albaladejo09b} pointing in the $x$ and $y$ 
directions (optic plane) polarized along $z$ with a phase difference $\theta = 
\pi/2\ \mathrm{rad}$ (see Fig.  \ref{fig1}).  This \textit{primary} field with 
wavelength $\lambda$ and wavenumber $k = 2 \pi / \lambda$ equals 
$\mathbf{E}_0(\mathbf{r}) = \ii 2 |E_0| \left(\sin(kx) + e^{\ii \theta} \sin(ky) 
\right) \hat{\bf{z}}$ with $|E_0|^2=2 I n/(c \epsilon\epsilon_0)$ controlled by 
the intensity of the laser beam $I$ and the refractive index of water $n = 
\sqrt{\epsilon}$ ($\epsilon$ is the relative permittivity and $c$ is the vacuum 
speed of light).  The electrodynamic response of spherical nanoparticles, much 
smaller than the laser wavelength $R \ll \lambda$, can be well described by 
their complex electrical polarizability $\alpha=\alpha' + \ii \alpha''$ (see 
Supplemental Material, SM). The  spatially   modulated  field
$\mathbf{E}_{exc}$    will   induce    a   dipole    $\mathbf{p}_i   =
\epsilon\epsilon_{0}      \alpha\mathbf{E}_{\text{exc}}(\mathbf{r}_i)$
centered  on  the  position   of  each  particle  $\mathbf{r}_i$ 
and, consequently, a  non-vanishing averaged Lorentz force  with components
${\bf  F}  (\textbf{r}_{i})  =  (\epsilon\epsilon_{0}/2)  \Re  \left\{
\alpha\left[\mathbf{E}_{\mathrm{exc}}(\mathbf{r})\cdot
  \nabla\mathbf{E}_{\mathrm{exc}}^*(\mathbf{r})\right]
\right\}_{\mathbf{r}  =  \mathbf{r}_i}$.
The primary  optical  force arises from $\mathbf{E}_{\text{exc}}=\mathbf{E}_{0}$,
\begin{align}
  \label{eq:force_field}
  \bfuno(x,y) = & 2 \alpha' \frac{n}{c} I\ \bm{\nabla}\left(\sin^2 (kx)
                                                            + \sin^2 (ky)\right)
  \nonumber \\
  & + 2 \alpha'' \frac{n}{c} I\
                 \bm{ \nabla} \times \left[2 \cos (kx)
                                           \cos (ky)\ \mathbf{e}_z\right].
\end{align}

This {\em nonconservative} force field \eqref{eq:force_field} forms a periodic 
pattern of optical vortices \cite{hemmerich1992radiation}, with a Bravais unit 
cell in $(x,y) \in [-\lambda/2,\lambda/2]$ (see Fig.  \ref{fig1}(b)).  Its 
non-conservative part $\nabla \times A(x,y)\mathbf{\hat{z}}$, proportional to 
$\alpha''$ \cite{Note1} embodies the so-called ``curl forces'' 
\cite{Albaladejo09b,Berry13,Berry15,Zapata2015}.  It resembles a chequerboard, 
with alternating squares of positive and negative vorticity. Curl-forces neatly 
convert the laser energy into work \cite{hemmerich1992radiation,Berry13,Berry15} 
and $U \equiv 2I(n/c) \alpha''= \epsilon\epsilon_0 |E_0|^2 \alpha''$ (related to 
the energy per NP) is a convenient unit to compare with the thermal counterpart 
$k_B T$.  The primary field has no stationary nodes where the NP can rest 
\cite{Zapata09}, but four unstable saddle nodes (SD) instead (see Fig.  
\ref{fig1}(b)).  As a consequence, an isolated gold NP experiences giant normal 
diffusion \cite{Albaladejo09a} with a diffusion coefficient $D_{op} \sim 
\lambda^2/\tau_{op} = u \,D_{th}$ which is larger than the thermal value 
$D_{th}$ for $u \equiv U/k_BT>1$.  This nondimensional laser energy $u$ can be 
increased up to $\sim 10^2$ before cavitation takes place \cite{Sivan,Seol}.

We investigate the \textit{collective behaviour} of a \textit{suspension} of 
gold NPs.  Aside from the primary force in Eq. \eqref{eq:force_field} and the 
thermal kicks, NPs interact optically through multiple scattering forces 
$\mathbf{F}^{(2)}(\left\{\mathbf{r}\right\})$, and also hydrodynamically, with a 
collective drag velocity $\mathbf{v}^{(c)}_{i}=\sum_j \mu_{ij}(r_{ij}) 
\mathbf{F}_j$ arising from the \textit{mutual} hydrodynamic mobility $\mu_{ij}$ 
dominated by the Oseen term $\mu_{ij}\sim 1/r_{ij}$.

\begin{figure}
  \includegraphics[scale=0.45]{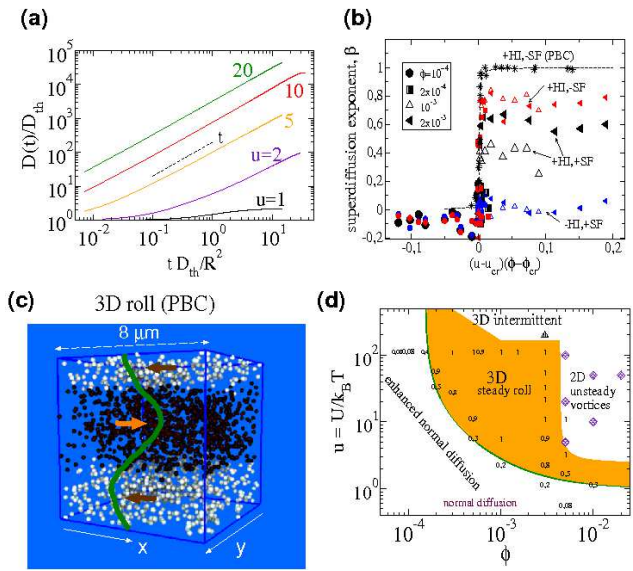}
  \caption{\label{fig2}
    (a) Effective diffusion coefficient of gold NPs $D(t)=\msdop/(4 t)$ in 
    periodic boundaries at $\phi = 2\times10^{-3}$, obtained from 
    single-particle in-plane MSD $\msdop(t)$ and scaled with the thermal value 
    $D_{th}=k_BT/(6\pi\eta R)$. For $U/K_BT=u>u_{cr} \sim 1$ the dynamics become 
    ballistic $D \sim t^{\beta}$ with $\beta = 1$.  (c) The superdiffusion 
    exponent $\beta$ against $\xi=[u-u_{cr}(\phi)][\phi-\phi_{cr}(u)]$ for 
    periodic (star symbols) and confined setups. Results with and without 
    hydrodynamic interactions ($+\mathrm{HI}$ and $-\mathrm{HI}$) and/or 
    secondary optical forces ($+\mathrm{SF}$ and $-\mathrm{SF}$) are compared. 
    (c) The coherent {\em 3D roll} of NPs recirculating in the periodic domain 
    with a sinusodal velocity profile in the $z$ direction. (d) Dynamic phase 
    diagram showing the transition from enhanced normal diffusion ($D \simeq u 
    D_{th}$) to coherent dynamics $D\sim t^{\beta}$ at the (green) critical line 
    $\xi^{*}=(u_{cr}-u^*)(\phi_{cr}-\phi^*)$ (see text for details).
  }
\end{figure}

In order to isolate the effect of hydrodynamic interactions (HI) at fixed 
$\phi$, we first solved these dynamics under periodic boundary conditions (PBC), 
without taking into account multiple scattering. These dynamics were solved 
using the Fluctuating Immersed Boundary method (FIB), which is an immersed 
boundary method for Stokesian particles in fluctuating hydrodynamics \cite{FIB} 
(see SM). We evaluate the (time-dependent) single particle diffusion defined as, 
$D(t) = \msdop/(4t)$ where $\msdop = \langle \left({\bf r}_1^{\parallel}(t_0 + 
t) -{\bf r}_1^{\parallel}(t_0)\right)^2\rangle$ is the in-plane $(\parallel)$ 
mean square displacement (MSD) of a tracer NP with position ${\bf r}_1= {\bf 
r}^{\parallel}_1 + z_1 {\bf {\hat e}}_z$. At a fixed $\phi$, Fig.  \ref{fig2}(a) 
shows a sudden transition from (enhanced) normal diffusion $D=D_{op}$ to 
superdiffusion $D(t)\sim t^{\beta}$, where the diffusion exponent jumps to the 
ballistic value $\beta=1$ above the critical laser energy $u>u_{cr}(\phi)$. In 
periodic boundaries, the superdiffusive regime corresponds to a large current of 
NPs occupying the whole box in the $z$ direction and flowing with vorticity in 
$\mathbf{d}= (\pm1, \pm1)$ direction. This {\em 3D roll}, illustrated in Fig. 
\ref{fig2}(c), would be qualitatively similar in an enclosure with walls.

The phase diagram of Fig. \ref{fig2}(d) draws the critical line 
$(u_{cr},\phi_{cr})$ separating uncorrelated walkers from the synchronized flow 
of NPs. Beyond, at larger $u$ or $\phi$, an extremely rich collective dynamics 
unfolds (more details in SM). The critical line satisfies $\xi^{*}=(u_{cr}-u^*) 
(\phi_{cr}-\phi^*)$ where $u^*$ and $\phi^*$ are threshold values below which 
the transition does not occur.  Under PBC we find $\xi^{*}=(1.55\pm 0.05)\times 
10^{-4}$ while $\phi^* = (1.5\pm 0.1)\time 10^{-4}$ and $u^* = 1.0$.  
Interestingly, coherent dynamics cannot ensue unless the energy per NP surpasses 
the thermal energy $U>U^* \simeq k_BT$.  Also remarkable, is the low threshold 
concentration $\phi^*$ involving an average separation of $3 \lambda$ between 
NPs.

To investigate the effect of multiple scattering we calculate the many-body 
secondary optic forces ${\bf F}_2({\bf r})$ (see SM) and plugged them into a 
Brownian solver implemented with the Rotne-Prager-Yamakawa mobility tensor for 
HI \cite{Ermak1978}.  To fix $\phi$ we first considered a confined domain where 
a repulsive external potential impedes NPs from escaping out from a cubic box of 
side $L = 7\lambda$. This confined geometry mimics an experimentally feasible 
laser-trap \cite{Dienerowitz}.  In this case we analyzed the collective 
diffusion evaluated from the in-plane displacement of the center of mass (CoM) 
of the NP ensemble $D_{cm}(t) = \msdcm(t)/(4 t)$ (details in the SM).  Under 
confinement, the onset of the superdiffusive regime $D_{cm}\sim t^{\beta}$ 
matches the findings under PBC, highlighting the hydrodynamic nature of the 
transition. Figure \ref{fig2}(b) plots values of $\beta$ (for both periodic and 
confined domains) against the governing non-dimensional group $\xi\equiv 
[u_{cr}(\phi)-u] [\phi_{cr}(u)-\phi]$. The transition is sharp and resembles the 
first-order dynamic transition reported for the Vicsek model \cite{Chate2012}. 
Confinement increases the value of the critical parameter to $\xi^* =\left(6\pm 
1\right) \times 10^{-4}$, but it does not alter the shape of the critical line 
nor significantly modify the threshold values ($u^* \simeq 0.7$ and $\phi \simeq 
(1.1 \pm 0.01)\times 10^{-4}$), suggesting a sort of ``material property'', as 
in the collective dynamics of micro-rollers \cite{bricard2013emergence}.

The ``microscopic'' origin of the coherent dynamics can be understood by ploting 
the single-NP probability density $\rho(\mathbf{r})$ and the streamlines of the 
NP average current, in Fig. \ref{fig1}(c). Above the critical energy [$u=2$ in 
Fig.\ref{fig1}(c)], superdiffusion arises as an instability of the basic 
stationary solution $\rho_0(\mathbf{r})$, illustrated in Fig.\ref{fig1}(b) 
$(u=1)$. The instability is triggered by the collective current 
$\mathbf{v}^{(c)}$ arising from HI. As shown in \ref{fig1}(b), this current 
breaks one of the symmetries of the primary optical field (mirrored by 
$\rho_0(\mathbf{r})$) and creates an average mass flow across the system. The 
particles tend to follow each other along zig-zag paths which move along one of 
the four possible diagonal directions of the lattice $\mathbf{d}= (\pm1, \pm1)$. 
This is consistent with hydrodynamic pairing \cite{sokolov2011hydrodynamic} 
which is known to happen if particles are forced to follow curved paths in a 
liquid. Note that the instability is degenerate in $\mathbf{d}$ and the NP 
current might jump from one diagonal to another, after many optical times 
$\tau_{op}$.

\begin{figure}
  \includegraphics[scale=0.5]{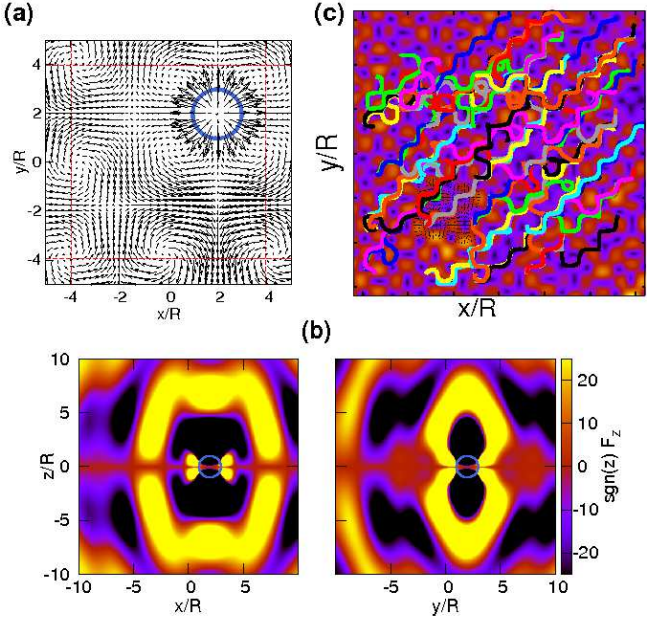}
  \caption{\label{fig3} (a)  The force field  on particle $i=2$,  in a
    system    with    two NPs    $\mathbf{F}_2=\mathbf{F}_2^{(1)}({\bf
    r}_2) +\mathbf{F}_2^{(2)}({\bf  r}_1, {\bf   r}_2$),  where  ${\bf
    F}^{(2)}_2$ comes from the light  scattered by NP $i = 1$, located
    at  $z=0$ and  plane location  $(1/4,1/4) \lambda$  (blue circle).
    The  dotted red  lines indicate  a separation  of $\lambda$.   (b)
    The corresponding normal  component of  the optical
    force $F_z$  in the $xz$  and $yz$ planes  times the sign  of $z$.
    Contours  saturate to  black  (attractive force,  $\mathrm{sgn}(z)
    F_z<0$) and yellow (repulsive) beyond  the values indicated in the
    colour scale.  Red contours, around  $z\simeq \pm\lambda = \pm 7.9
    \,R$ correspond to ``stable'' domains ($F_z = 0$ and $dF_z(z)/dz <
    0$).  Panel  (c) overlaps the  trajectories of several  NPs moving
    around  the  $z=0$  plane  of  a  confined  domain  ($L \simeq 2.9\  \mu\mathrm{m}$,
    $\phi=3\times 10^{-3}$ and  $u=2$) over  the isocontours  of the  total electric
    field       intensity       at        the       initial       time
    $|E(\mathbf{r},t_0)|$.  Scattering forces  induce disruptions  of
    the  zig-zag paths  and dislocations  of the  $\mathbf{E}$ lattice
    pattern. }
\end{figure}

Although the dynamic transition stem chiefly from hydrodynamic interactions (HI) 
[see case $-HI$ in Fig. \ref{fig2}(b)], secondary forces have a measurable 
effect in reducing the intensity of the collective current.  Secondary forces 
may be viewed as a self-generated, speckle intensity pattern induced by multiple 
scattering of light. This pattern disrupts the periodic stationary curl force 
field, thus reducing the coherence of the zig-zag paths followed by the NP's 
under coherent motion [see Fig \ref{fig3}(c)]. The phase lag term 
$\exp\left[\mathrm{i}\,\mathbf{k} \cdot \mathbf{r}_{ij}\right]$ present in the 
optical Green function (see SM) creates undulating force patterns 
\cite{Brugger15} which become quite complex, even for just two isolated NPs 
$\bfdos = \bfdos(\mathbf{r}_1,\mathbf{r}_2)$ (see Fig.\ref{fig3}).  
As illustrated in Fig.  \ref{fig3}(a), two NPs in the same optic plane tend to 
repel each other. In a confined domain (see SM), this leads to the expansion of 
the NP's ensemble volume.  However, along $z$, the scatter produces marginally 
stable domains [$F_z=0$ and $dF_z/dz<0$, see Fig.  \ref{fig3}(b)]. In 
suspension, this effect leads to transient layers of NPs separated by roughly 
$\Delta z \sim 2 \lambda$ (SM).

\begin{figure}
  \includegraphics[scale=0.4,angle=-180]{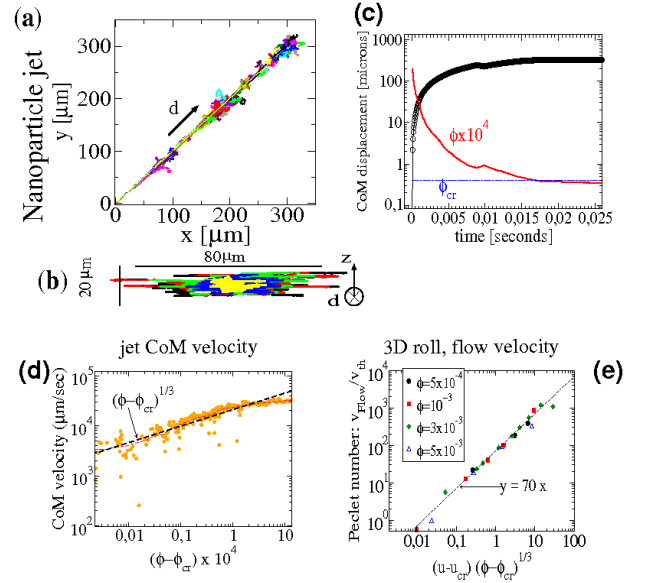}
  \caption{\label{fig4} Nanoparticle jet:  (a)  Trajectories in the
    optic plane of  80 nanoparticles after being released from their confinement in a
    box of  $L \simeq 2.9\  \mu\mathrm{m}$;  the jet moves in direction $\mathbf{d}  =    \hat{\mathbf{x}} + \hat{\mathbf{y}}$ (diferent colors for  different NPs).
    (b) Dispersion of the NP's in the $zd$  plane (colors superpose for increasing times).
    (c) Displacement  of the center of  mass (CoM) and (c) the CoM  velocity $v_{cm}$ versus the
    instantaneous   volume  fraction   $\phi(t)$  (measured   from  the  NP's ensemble volume
    $\mathcal{V}=\prod_{\alpha} R_{\alpha}^{1/2}$, where $\mathbf{R}$ are the eigenvalues
    of the ensemble's gyration tensor).  The scaling $v_{cm} \sim (\phi-\phi_{cr})^{1/3}$.
    showing the  peak of the velocity  profile of the 3D  roll NP-flow
    (see Fig.  \ref{fig2}(c)) in a periodic  domain $v_{\mathrm{Flow}}
    \propto (u-u_{cr})\,(\phi-\phi_{cr})^{1/3}$
  }
\end{figure}

A nontrivial question  is what happens if the  confinement is removed.
As a possible outcome, the  extra pressure from secondary forces might
destroy coherent motion by inducing a fast dispersion of NPs over the
optic  plane. We  prepared  experiments placing  NPs  in the  confined
domain,  with  $u$ and  initial  concentration  $\phi(t=0)$ above  the
transition line. Upon removing the confining potential, we
immediately  observe  the formation  of  a  jet  of  NPs moving  at an
impressive velocity (see video in SM).  The trajectories  of single NPs, the CoM
displacement and  velocity $v_{cm}(t)$  are shown in  Fig.  \ref{fig4}
for one of the cases considered. The jet moves along one the diagonals
${\bf d}$  and its direction  remains stable until  dispersion (mostly
along ${\bf  d}$ direction) leads to  $\phi(t)<\phi_{cr}(u)$       [see
  Fig. \ref{fig4}(b)]. Notably,  the flock disperses much  less in $z$
direction  when  secondary  forces  are added. As shown in SM,  attractive
secondary normal forces counter-balance  the  repulsive
Oseen drag in  $z$ direction.
As shown  in Fig. \ref{fig4}(c) and  (d), the collective
velocity scales  like $v_{cm}\propto  (\phi -  \phi_{cr})^{1/3}$ which
differs  from  that  observed  in the collective  motion  of  microrollers
$v_{cm}                     \sim                     (\phi-\phi_{cr})$
\cite{bricard2013emergence,aleksNature}.   A  simple scaling  argument
unifies  both  results.   In   terms  of  the  interparticle  distance
$r_{12}$, the  leading term  in the  mutual mobility  generally scales
like  $\mu_{12}\sim  1/r_{12}^{\alpha}$  while, in  a  $d$-dimensional
system,  $r_{12} \sim  \phi^{-1/d}$. One  thus expects  the collective
velocity  $v_{cm}  \sim  \mu_{12}  F_2$ to  scale  like  $v_{cm}  \sim
\phi^{\alpha/d}\, F$.   Groups of microrollers move  with $v_{cm} \sim
\phi$ \cite{bricard2013emergence,aleksNature}  close to a  wall, where
$d = 2$ and $\alpha = 2$. In our  setup $d = 3$ and the laser force $F
\sim  U/\lambda$ activates  monopole (Oseen)  couplings $\alpha  = 1$,
providing $v_{cm} \sim \phi^{1/3} U/\lambda$. To complete our scaling,
we impose  $v_{cm} = 0$  below the  critical line, leading  to $v_{cm}
\propto   \left(u-u_{cr}\right)  \left(\phi   -\phi_{cr}\right)^{1/3}$
which  agrees  quite well  with  results  in periodic, unconfined and confined
domains, as shown respectively in (Fig. \ref{fig4}(c), (d) and SM (confined case).

We have presented a collective  dynamic transition of optically driven
gold   nanoparticles,   which   shares  many   generic   features   of
(deterministic)      micron-sized     active      colloidal     flocks
\cite{bricard2013emergence,aleksNature}.  By constrast,  here, thermal
forces  are  strong  but  also essential for the enhanced diffusion and
the collective dynamics.  Without  fluctuations the  NPs would  frezee at  the saddle
nodes. Calculations under extreme energies $u>200$,
show that NPs take then too long  to leave
the SD domains,  leading to intermitency and  supression of collective
motion (see Fig.  \ref{fig2}(d) and SM).  This is strongly reminiscent
of \textit{  stochastic resonance}  \cite{SR} where fluctuations
trigger coherent dynamics within a  finite window of noise amplitudes.
Here,  stochastic  forces  ``excite''  NP
displacements in a \textit{stationary optical field} leading to strong
(monopolar)  hydrodynamics driving  jets of  \textit{nanoparticles} at
velocities  of  $\mathrm{cm/s}$,  similar  to  micro-roller
clusters  \cite{aleksNature}, but orders of magnitude larger than recent
optofluidic setups based on deterministic \cite{nature2016} or stochastic driving \cite{Douglass12,volpe14}. The  required  power   densities  are
standard for   optical    tweezers    \cite{Brugger15} $I   \sim 10^{9}\ \mathrm{W/m}^2$,
attained by focusing a $0.1\ \mathrm{W}$ laser onto  a $\sim  10\ \mu\mathrm{m}$-side  domain. Close  to the  plasmon
resonance,   local   heating   is   expected   upon   increasing   $I$
\cite{Seol,Yeshchenko} so these dynamics might  then provide a new way
for    ultrafast    transport    of     heat    at    the    nanoscale
\cite{baffou2013thermo}.   Also, in  our  setup the  nanoscopic-length
Peclet  number reaches $\mathrm{Pe} = v_{cm}  R \rho_f  /\eta \sim 10^2$,
promising   a  novel   route   for  rapid   mixing  in   microdroplets
\cite{mix_nano_rods}.

\section*{Acknowledgements}
We thank Florencio Balboa for his help in the PBC calculations and Raul P\'erez for help in RPY implementation. This  research  was  supported  by   the  Spanish  Ministerio  de  Econom\'ia  y Competitividad (MICINN)  and European  Regional Development Fund  (ERDF) through Projects Explora NANOMIX FIS2013-50510-EXP,  FIS2015-69295-C3-3-P, MDM-2014-0377 and the Basque
Dep.\ de Educaci\'on through Project PI-2016-1-0041 (J.J.S.).
R.D-B acknowledges ACS-PRF grant number 54312-ND9.

\bibliography{Nanomix.v7}

\begin{thebibliography}{30}
\expandafter\ifx\csname natexlab\endcsname\relax\def\natexlab#1{#1}\fi
\expandafter\ifx\csname bibnamefont\endcsname\relax
  \def\bibnamefont#1{#1}\fi
\expandafter\ifx\csname bibfnamefont\endcsname\relax
  \def\bibfnamefont#1{#1}\fi
\expandafter\ifx\csname citenamefont\endcsname\relax
  \def\citenamefont#1{#1}\fi
\expandafter\ifx\csname url\endcsname\relax
  \def\url#1{\texttt{#1}}\fi
\expandafter\ifx\csname urlprefix\endcsname\relax\def\urlprefix{URL }\fi
\providecommand{\bibinfo}[2]{#2}
\providecommand{\eprint}[2][]{\url{#2}}

\bibitem[{\citenamefont{Ndukaife et~al.}(2016)\citenamefont{Ndukaife,
  Kildishev, Nannna, Shalaev, Wereley, and Boltasseva}}]{nature2016}
\bibinfo{author}{\bibfnamefont{J.~C.} \bibnamefont{Ndukaife}},
  \bibinfo{author}{\bibfnamefont{A.~V.} \bibnamefont{Kildishev}},
  \bibinfo{author}{\bibfnamefont{A.~G.~A.} \bibnamefont{Nannna}},
  \bibinfo{author}{\bibfnamefont{V.~M.} \bibnamefont{Shalaev}},
  \bibinfo{author}{\bibfnamefont{S.~T.} \bibnamefont{Wereley}},
  \bibnamefont{and}
  \bibinfo{author}{\bibfnamefont{A.}~\bibnamefont{Boltasseva}},
  \bibinfo{journal}{Nature Nanotechnology} \textbf{\bibinfo{volume}{11}},
  \bibinfo{pages}{53} (\bibinfo{year}{2016}).

\bibitem[{\citenamefont{{J}ones et~al.}(2015)\citenamefont{{J}ones, {M}arag\'o,
  and {V}olpe}}]{Jones}
\bibinfo{author}{\bibfnamefont{P.~H.} \bibnamefont{{J}ones}},
  \bibinfo{author}{\bibfnamefont{O.~M.} \bibnamefont{{M}arag\'o}},
  \bibnamefont{and} \bibinfo{author}{\bibfnamefont{G.}~\bibnamefont{{V}olpe}},
  \emph{\bibinfo{title}{{O}ptical {T}weezers: {P}rinciples and {A}ppilactions}}
  (\bibinfo{publisher}{{C}ambridge {U}niversity {P}ress},
  \bibinfo{address}{{C}ambridge}, \bibinfo{year}{2015}).

\bibitem[{\citenamefont{{D}ouglass et~al.}(2012)\citenamefont{{D}ouglass,
  {S}ukhov, and {D}ogariu}}]{Douglass12}
\bibinfo{author}{\bibfnamefont{K.~M.} \bibnamefont{{D}ouglass}},
  \bibinfo{author}{\bibfnamefont{S.}~\bibnamefont{{S}ukhov}}, \bibnamefont{and}
  \bibinfo{author}{\bibfnamefont{A.}~\bibnamefont{{D}ogariu}},
  \bibinfo{journal}{{N}ature {P}hotonics} \textbf{\bibinfo{volume}{6}},
  \bibinfo{pages}{834} (\bibinfo{year}{2012}).

\bibitem[{\citenamefont{Volpe et~al.}(2014)\citenamefont{Volpe, Volpe, and
  Gigan}}]{volpe14}
\bibinfo{author}{\bibfnamefont{G.}~\bibnamefont{Volpe}},
  \bibinfo{author}{\bibfnamefont{G.}~\bibnamefont{Volpe}}, \bibnamefont{and}
  \bibinfo{author}{\bibfnamefont{S.}~\bibnamefont{Gigan}},
  \bibinfo{journal}{Scientific Reports} \textbf{\bibinfo{volume}{4}},
  \bibinfo{pages}{3936} (\bibinfo{year}{2014}).

\bibitem[{\citenamefont{{B}r\"ugger et~al.}(2015)\citenamefont{{B}r\"ugger,
  {F}roufe {P}\'erez, {S}cheffold, and {S}\'aenz}}]{Brugger15}
\bibinfo{author}{\bibfnamefont{G.}~\bibnamefont{{B}r\"ugger}},
  \bibinfo{author}{\bibfnamefont{L.~S.} \bibnamefont{{F}roufe {P}\'erez}},
  \bibinfo{author}{\bibfnamefont{F.}~\bibnamefont{{S}cheffold}},
  \bibnamefont{and} \bibinfo{author}{\bibfnamefont{J.~J.}
  \bibnamefont{{S}\'aenz}}, \bibinfo{journal}{{N}ature {C}ommunications}
  \textbf{\bibinfo{volume}{6}}, \bibinfo{pages}{7460} (\bibinfo{year}{2015}).

\bibitem[{\citenamefont{Bechinger et~al.}(2016)\citenamefont{Bechinger,
  Di~Leonardo, L{\"o}wen, Reichhardt, Volpe, and Volpe}}]{bechinger2016active}
\bibinfo{author}{\bibfnamefont{C.}~\bibnamefont{Bechinger}},
  \bibinfo{author}{\bibfnamefont{R.}~\bibnamefont{Di~Leonardo}},
  \bibinfo{author}{\bibfnamefont{H.}~\bibnamefont{L{\"o}wen}},
  \bibinfo{author}{\bibfnamefont{C.}~\bibnamefont{Reichhardt}},
  \bibinfo{author}{\bibfnamefont{G.}~\bibnamefont{Volpe}}, \bibnamefont{and}
  \bibinfo{author}{\bibfnamefont{G.}~\bibnamefont{Volpe}},
  \bibinfo{journal}{Reviews of Modern Physics} \textbf{\bibinfo{volume}{88}},
  \bibinfo{pages}{045006} (\bibinfo{year}{2016}).

\bibitem[{\citenamefont{Vicsek and Zafeiris}(2012)}]{Vicsek2012}
\bibinfo{author}{\bibfnamefont{T.}~\bibnamefont{Vicsek}} \bibnamefont{and}
  \bibinfo{author}{\bibfnamefont{A.}~\bibnamefont{Zafeiris}},
  \bibinfo{journal}{Physics Reports} \textbf{\bibinfo{volume}{517}},
  \bibinfo{pages}{71} (\bibinfo{year}{2012}).

\bibitem[{\citenamefont{Driscoll et~al.}(2017)\citenamefont{Driscoll, Delmotte,
  Youssef, Scanna, Deonve, and Chaikin}}]{aleksNature}
\bibinfo{author}{\bibfnamefont{M.}~\bibnamefont{Driscoll}},
  \bibinfo{author}{\bibfnamefont{B.}~\bibnamefont{Delmotte}},
  \bibinfo{author}{\bibfnamefont{M.}~\bibnamefont{Youssef}},
  \bibinfo{author}{\bibfnamefont{S.}~\bibnamefont{Scanna}},
  \bibinfo{author}{\bibfnamefont{A.}~\bibnamefont{Deonve}}, \bibnamefont{and}
  \bibinfo{author}{\bibfnamefont{P.}~\bibnamefont{Chaikin}},
  \bibinfo{journal}{Nature Physics} \textbf{\bibinfo{volume}{13}},
  \bibinfo{pages}{375} (\bibinfo{year}{2017}).

\bibitem[{\citenamefont{Bricard et~al.}(2013)\citenamefont{Bricard, Caussin,
  Desreumaux, Dauchot, and Bartolo}}]{bricard2013emergence}
\bibinfo{author}{\bibfnamefont{A.}~\bibnamefont{Bricard}},
  \bibinfo{author}{\bibfnamefont{J.-B.} \bibnamefont{Caussin}},
  \bibinfo{author}{\bibfnamefont{N.}~\bibnamefont{Desreumaux}},
  \bibinfo{author}{\bibfnamefont{O.}~\bibnamefont{Dauchot}}, \bibnamefont{and}
  \bibinfo{author}{\bibfnamefont{D.}~\bibnamefont{Bartolo}},
  \bibinfo{journal}{Nature} \textbf{\bibinfo{volume}{503}}, \bibinfo{pages}{95}
  (\bibinfo{year}{2013}).

\bibitem[{\citenamefont{Theraulaz et~al.}(2002)\citenamefont{Theraulaz,
  Bonabeau, Nicolis, Sol\'e, Fourcassi\'e, Blanco, Fournier, Joly, Fern\'andez,
  Grimal et~al.}}]{ants}
\bibinfo{author}{\bibfnamefont{G.}~\bibnamefont{Theraulaz}},
  \bibinfo{author}{\bibfnamefont{E.}~\bibnamefont{Bonabeau}},
  \bibinfo{author}{\bibfnamefont{S.~C.} \bibnamefont{Nicolis}},
  \bibinfo{author}{\bibfnamefont{R.~V.} \bibnamefont{Sol\'e}},
  \bibinfo{author}{\bibfnamefont{V.}~\bibnamefont{Fourcassi\'e}},
  \bibinfo{author}{\bibfnamefont{S.}~\bibnamefont{Blanco}},
  \bibinfo{author}{\bibfnamefont{R.}~\bibnamefont{Fournier}},
  \bibinfo{author}{\bibfnamefont{J.-L.} \bibnamefont{Joly}},
  \bibinfo{author}{\bibfnamefont{P.}~\bibnamefont{Fern\'andez}},
  \bibinfo{author}{\bibfnamefont{A.}~\bibnamefont{Grimal}},
  \bibnamefont{et~al.}, \bibinfo{journal}{PNAS} \textbf{\bibinfo{volume}{15}},
  \bibinfo{pages}{9645} (\bibinfo{year}{2002}),
  \urlprefix\url{http://europepmc.org/articles/PMC124961}.

\bibitem[{\citenamefont{Brzobohat{\`y}
  et~al.}(2015)\citenamefont{Brzobohat{\`y}, Arzola, {\v{S}}iler, Chv{\'a}tal,
  J{\'a}kl, Simpson, and Zem{\'a}nek}}]{brzobohaty2015complex}
\bibinfo{author}{\bibfnamefont{O.}~\bibnamefont{Brzobohat{\`y}}},
  \bibinfo{author}{\bibfnamefont{A.~V.} \bibnamefont{Arzola}},
  \bibinfo{author}{\bibfnamefont{M.}~\bibnamefont{{\v{S}}iler}},
  \bibinfo{author}{\bibfnamefont{L.}~\bibnamefont{Chv{\'a}tal}},
  \bibinfo{author}{\bibfnamefont{P.}~\bibnamefont{J{\'a}kl}},
  \bibinfo{author}{\bibfnamefont{S.}~\bibnamefont{Simpson}}, \bibnamefont{and}
  \bibinfo{author}{\bibfnamefont{P.}~\bibnamefont{Zem{\'a}nek}},
  \bibinfo{journal}{Optics Express} \textbf{\bibinfo{volume}{23}},
  \bibinfo{pages}{7273} (\bibinfo{year}{2015}).

\bibitem[{\citenamefont{Hemmerich and
  H{\"a}nsch}(1992)}]{hemmerich1992radiation}
\bibinfo{author}{\bibfnamefont{A.}~\bibnamefont{Hemmerich}} \bibnamefont{and}
  \bibinfo{author}{\bibfnamefont{T.}~\bibnamefont{H{\"a}nsch}},
  \bibinfo{journal}{Physical review letters} \textbf{\bibinfo{volume}{68}},
  \bibinfo{pages}{1492} (\bibinfo{year}{1992}).

\bibitem[{\citenamefont{{A}lbaladejo
  et~al.}(2009{\natexlab{a}})\citenamefont{{A}lbaladejo, {M}arqu\'es,
  {L}aroche, and {S}\'aenz}}]{Albaladejo09b}
\bibinfo{author}{\bibfnamefont{S.}~\bibnamefont{{A}lbaladejo}},
  \bibinfo{author}{\bibfnamefont{M.~I.} \bibnamefont{{M}arqu\'es}},
  \bibinfo{author}{\bibfnamefont{M.}~\bibnamefont{{L}aroche}},
  \bibnamefont{and} \bibinfo{author}{\bibfnamefont{J.~J.}
  \bibnamefont{{S}\'aenz}}, \bibinfo{journal}{{P}hysical {R}eview {L}etters}
  \textbf{\bibinfo{volume}{102}}, \bibinfo{pages}{113602}
  (\bibinfo{year}{2009}{\natexlab{a}}).

\bibitem[{\citenamefont{{A}lbaladejo
  et~al.}(2009{\natexlab{b}})\citenamefont{{A}lbaladejo, {M}arqu\'es,
  {S}cheffold, and {S}\'aenz}}]{Albaladejo09a}
\bibinfo{author}{\bibfnamefont{S.}~\bibnamefont{{A}lbaladejo}},
  \bibinfo{author}{\bibfnamefont{M.~I.} \bibnamefont{{M}arqu\'es}},
  \bibinfo{author}{\bibfnamefont{F.}~\bibnamefont{{S}cheffold}},
  \bibnamefont{and} \bibinfo{author}{\bibfnamefont{J.~J.}
  \bibnamefont{{S}\'aenz}}, \bibinfo{journal}{{N}anoletters}
  \textbf{\bibinfo{volume}{9}}, \bibinfo{pages}{3527}
  (\bibinfo{year}{2009}{\natexlab{b}}).

\bibitem[{\citenamefont{{Z}apata et~al.}(2009)\citenamefont{{Z}apata,
  {A}lbaladejo, {P}arrondo, {S}\'aenz, and {S}ols}}]{Zapata09}
\bibinfo{author}{\bibfnamefont{I.}~\bibnamefont{{Z}apata}},
  \bibinfo{author}{\bibfnamefont{S.}~\bibnamefont{{A}lbaladejo}},
  \bibinfo{author}{\bibfnamefont{J.~M.~R.} \bibnamefont{{P}arrondo}},
  \bibinfo{author}{\bibfnamefont{J.~J.} \bibnamefont{{S}\'aenz}},
  \bibnamefont{and} \bibinfo{author}{\bibfnamefont{F.}~\bibnamefont{{S}ols}},
  \bibinfo{journal}{{P}hysical {R}eview {L}etters}
  \textbf{\bibinfo{volume}{103}}, \bibinfo{pages}{130601}
  (\bibinfo{year}{2009}).

\bibitem[{\citenamefont{{B}erry and {S}hukla}(2013)}]{Berry13}
\bibinfo{author}{\bibfnamefont{M.~V.} \bibnamefont{{B}erry}} \bibnamefont{and}
  \bibinfo{author}{\bibfnamefont{P.}~\bibnamefont{{S}hukla}},
  \bibinfo{journal}{{J}ournal of {P}hysics {A}: {M}athematical and
  {T}heoretical} \textbf{\bibinfo{volume}{46}}, \bibinfo{pages}{422001}
  (\bibinfo{year}{2013}).

\bibitem[{\citenamefont{{B}erry and {S}hukla}(2015)}]{Berry15}
\bibinfo{author}{\bibfnamefont{M.~V.} \bibnamefont{{B}erry}} \bibnamefont{and}
  \bibinfo{author}{\bibfnamefont{P.}~\bibnamefont{{S}hukla}},
  \bibinfo{journal}{{P}roceedings of the {R}oyal {S}ociety of {L}ondon {A}:
  {M}athematical, {P}hysical and {E}ngineering {S}ciences}
  \textbf{\bibinfo{volume}{471}} (\bibinfo{year}{2015}).

\bibitem[{\citenamefont{Zapata et~al.}(2016)\citenamefont{Zapata,
  Delgado-Buscalioni, and S{\'a}enz}}]{Zapata2015}
\bibinfo{author}{\bibfnamefont{I.}~\bibnamefont{Zapata}},
  \bibinfo{author}{\bibfnamefont{R.}~\bibnamefont{Delgado-Buscalioni}},
  \bibnamefont{and} \bibinfo{author}{\bibfnamefont{J.~J.}
  \bibnamefont{S{\'a}enz}}, \bibinfo{journal}{Phys. Rev. E}
  \textbf{\bibinfo{volume}{93}}, \bibinfo{pages}{0621130}
  (\bibinfo{year}{2016}).

\bibitem[{Not()}]{Note1}
\bibinfo{note}{To maximize the ratio $|\alpha ''|/|\alpha '|$ we consider a
  wavelength close to the gold plasmon resonance in water $\lambda \approx 395\
  \protect \mathrm {nm}$, setting $\epsilon =1.8$, $\alpha ' \approx 1\times
  10^{-21}\ \protect \mathrm {m}^3$ and $\alpha '' \approx 2\alpha '$}.

\bibitem[{\citenamefont{Sivan and Chu}(2017)}]{Sivan}
\bibinfo{author}{\bibfnamefont{Y.}~\bibnamefont{Sivan}} \bibnamefont{and}
  \bibinfo{author}{\bibfnamefont{S.-W.} \bibnamefont{Chu}},
  \bibinfo{journal}{Nanophotonics} \textbf{\bibinfo{volume}{6}},
  \bibinfo{pages}{317} (\bibinfo{year}{2017}).

\bibitem[{\citenamefont{Seol et~al.}(2006)\citenamefont{Seol, Carpenter, and
  Perkins}}]{Seol}
\bibinfo{author}{\bibfnamefont{Y.}~\bibnamefont{Seol}},
  \bibinfo{author}{\bibfnamefont{A.~E.} \bibnamefont{Carpenter}},
  \bibnamefont{and} \bibinfo{author}{\bibfnamefont{T.~T.}
  \bibnamefont{Perkins}}, \bibinfo{journal}{Optics letters}
  \textbf{\bibinfo{volume}{31}}, \bibinfo{pages}{2429} (\bibinfo{year}{2006}).

\bibitem[{\citenamefont{Delong et~al.}(2014)\citenamefont{Delong, Usabiaga,
  Delgado-Buscalioni, Griffith, and Donev}}]{FIB}
\bibinfo{author}{\bibfnamefont{S.}~\bibnamefont{Delong}},
  \bibinfo{author}{\bibfnamefont{F.~B.} \bibnamefont{Usabiaga}},
  \bibinfo{author}{\bibfnamefont{R.}~\bibnamefont{Delgado-Buscalioni}},
  \bibinfo{author}{\bibfnamefont{B.~E.} \bibnamefont{Griffith}},
  \bibnamefont{and} \bibinfo{author}{\bibfnamefont{A.}~\bibnamefont{Donev}},
  \bibinfo{journal}{The Journal of chemical physics}
  \textbf{\bibinfo{volume}{140}}, \bibinfo{pages}{134110}
  (\bibinfo{year}{2014}).

\bibitem[{\citenamefont{Ermak and McCammon}(1978)}]{Ermak1978}
\bibinfo{author}{\bibfnamefont{D.~L.} \bibnamefont{Ermak}} \bibnamefont{and}
  \bibinfo{author}{\bibfnamefont{J.}~\bibnamefont{McCammon}},
  \bibinfo{journal}{J. Chem. Phys.} \textbf{\bibinfo{volume}{69}},
  \bibinfo{pages}{1352} (\bibinfo{year}{1978}).

\bibitem[{\citenamefont{Dienerowitz et~al.}(2008)\citenamefont{Dienerowitz,
  Mazilu, Reece, Krauss, and Dholakia}}]{Dienerowitz}
\bibinfo{author}{\bibfnamefont{M.}~\bibnamefont{Dienerowitz}},
  \bibinfo{author}{\bibfnamefont{M.}~\bibnamefont{Mazilu}},
  \bibinfo{author}{\bibfnamefont{P.~J.} \bibnamefont{Reece}},
  \bibinfo{author}{\bibfnamefont{T.~F.} \bibnamefont{Krauss}},
  \bibnamefont{and} \bibinfo{author}{\bibfnamefont{K.}~\bibnamefont{Dholakia}},
  \bibinfo{journal}{Optics Express} \textbf{\bibinfo{volume}{16}},
  \bibinfo{pages}{4991} (\bibinfo{year}{2008}).

\bibitem[{\citenamefont{Gregoire and Chate}(2012)}]{Chate2012}
\bibinfo{author}{\bibfnamefont{G.}~\bibnamefont{Gregoire}} \bibnamefont{and}
  \bibinfo{author}{\bibfnamefont{H.}~\bibnamefont{Chate}},
  \bibinfo{journal}{Phys. Rev. Lett.} \textbf{\bibinfo{volume}{92}},
  \bibinfo{pages}{025702} (\bibinfo{year}{2012}).

\bibitem[{\citenamefont{Sokolov et~al.}(2011)\citenamefont{Sokolov, Frydel,
  Grier, Diamant, and Roichman}}]{sokolov2011hydrodynamic}
\bibinfo{author}{\bibfnamefont{Y.}~\bibnamefont{Sokolov}},
  \bibinfo{author}{\bibfnamefont{D.}~\bibnamefont{Frydel}},
  \bibinfo{author}{\bibfnamefont{D.~G.} \bibnamefont{Grier}},
  \bibinfo{author}{\bibfnamefont{H.}~\bibnamefont{Diamant}}, \bibnamefont{and}
  \bibinfo{author}{\bibfnamefont{Y.}~\bibnamefont{Roichman}},
  \bibinfo{journal}{Physical review letters} \textbf{\bibinfo{volume}{107}},
  \bibinfo{pages}{158302} (\bibinfo{year}{2011}).

\bibitem[{\citenamefont{Gammaitoni et~al.}(1998)\citenamefont{Gammaitoni,
  H{\"a}nggi, Jung, and Marchesoni}}]{SR}
\bibinfo{author}{\bibfnamefont{L.}~\bibnamefont{Gammaitoni}},
  \bibinfo{author}{\bibfnamefont{P.}~\bibnamefont{H{\"a}nggi}},
  \bibinfo{author}{\bibfnamefont{P.}~\bibnamefont{Jung}}, \bibnamefont{and}
  \bibinfo{author}{\bibfnamefont{F.}~\bibnamefont{Marchesoni}},
  \bibinfo{journal}{Reviews of Modern Physics} \textbf{\bibinfo{volume}{70}},
  \bibinfo{pages}{223} (\bibinfo{year}{1998}).

\bibitem[{\citenamefont{Yeshchenko et~al.}(2016)\citenamefont{Yeshchenko,
  Kutsevol, and Naumenko}}]{Yeshchenko}
\bibinfo{author}{\bibfnamefont{O.~A.} \bibnamefont{Yeshchenko}},
  \bibinfo{author}{\bibfnamefont{N.~V.} \bibnamefont{Kutsevol}},
  \bibnamefont{and} \bibinfo{author}{\bibfnamefont{A.~P.}
  \bibnamefont{Naumenko}}, \bibinfo{journal}{Plasmonics}
  \textbf{\bibinfo{volume}{11}}, \bibinfo{pages}{345} (\bibinfo{year}{2016}).

\bibitem[{\citenamefont{Baffou and Quidant}(2013)}]{baffou2013thermo}
\bibinfo{author}{\bibfnamefont{G.}~\bibnamefont{Baffou}} \bibnamefont{and}
  \bibinfo{author}{\bibfnamefont{R.}~\bibnamefont{Quidant}},
  \bibinfo{journal}{Laser \& Photonics Reviews} \textbf{\bibinfo{volume}{7}},
  \bibinfo{pages}{171} (\bibinfo{year}{2013}).

\bibitem[{\citenamefont{Chong et~al.}(2013)\citenamefont{Chong, Chin, Tan,
  Wang, Liu, and Chen}}]{mix_nano_rods}
\bibinfo{author}{\bibfnamefont{W.~H.} \bibnamefont{Chong}},
  \bibinfo{author}{\bibfnamefont{L.~K.} \bibnamefont{Chin}},
  \bibinfo{author}{\bibfnamefont{R.~L.~S.} \bibnamefont{Tan}},
  \bibinfo{author}{\bibfnamefont{H.}~\bibnamefont{Wang}},
  \bibinfo{author}{\bibfnamefont{A.~Q.} \bibnamefont{Liu}}, \bibnamefont{and}
  \bibinfo{author}{\bibfnamefont{H.}~\bibnamefont{Chen}},
  \bibinfo{journal}{Angew. Chem. Int. Ed.} \textbf{\bibinfo{volume}{52}},
  \bibinfo{pages}{8750} (\bibinfo{year}{2013}).

\end{thebibliography}

\end{document}